\input{aipcheck.tex}

\newcommand\sps{\space\space\space\space}
\documentclass[
  ]
  {aipproc}

\layoutstyle{8x11single}

\SetInternalRegister\hbadness{8000} % pseudo latin isn't breaking very well :-)

\newcommand\doingARLO[2][]{%
  \ifx\mmref\undefined #1\else #2\fi
}

\begin{document}

\title 
      [Blazar Demographics with MOJAVE and GLAST]
      {Blazar Demographics with MOJAVE and GLAST}


\author{Matthew L. Lister}{
  address={Department of Physics, Purdue University, West Lafayette, IN 47907-2036},
  email={mlister@physics.purdue.edu},
%  thanks={This work was commissioned by the AIP}
}

\iftrue
%\author{Mihai Cara}{
%  address={Department of Physics, Purdue University, West Lafayette, IN 47907-2036},
%  email={mcara@physics.purdue.edu},
%}

\fi

\copyrightyear  {2007}

\begin{abstract}

 MOJAVE is a long term VLBA program to investigate the kinematics and
 polarization evolution of a complete sample of 133 active galactic
 nuclei selected on the basis of compact, relativistically beamed jet
 emission at 15 GHz.  By fitting to the apparent distributions of
 superluminal speed and jet luminosity, we can constrain the Lorentz
 factor distribution and intrinsic luminosity function of the
 radio-selected blazar parent population. These low-energy peaked
 blazars formed a significant fraction of all EGRET detections, and
 should figure prominently in the GLAST source catalog. Using simple
 models, we investigate the predicted distribution of GLAST blazars in
 the gamma-ray/radio flux density plane, and describe an extension of
 the MOJAVE survey that will provide extensive parsec-scale jet
 information in complete regions of this plane. We find that if a
 population of intrinsically radio bright yet gamma-ray weak blazars
 exists, its signal will be largely wiped out by the large gamma-ray
 flux scatter associated with Doppler beaming.

\end{abstract}

\date{\today}

\maketitle

\section{Monte Carlo Blazar Population Simulations}

%\ifthenelse{\equal\selectedlayoutstyle{6x9}}{\par\bfseries 
\ifthenelse{\equal{8x11single}{6x9}}{\par\bfseries 
  Note:t}{}

Many of the overall characteristics of blazars and their parent
population are poorly known, mainly because of the inherent
difficulties in disentangling beaming and selection effects from
flux-limited samples. These issues can be addressed, however, given a
general knowledge of the distribution of bulk Lorentz and Doppler
factors in the blazar parent population. Since 1994 we have been
carrying out a large VLBA survey to determine these
quantities for the brightest AGN jets in the sky at a wavelength of 2
cm
\citep{KL04}. In 2002, full polarization imaging was added as part of the
MOJAVE (Monitoring Of Jets in Active galactic nuclei with VLBA
Experiments; http://www.physics.purdue.edu/astro/MOJAVE) program
\citep{LH05}, and the sample was refined to include all AGN with
J2000 declination $>$ $-$20$^\circ$, galactic latitude $|b| >
2.5^\circ$, and VLBA 2 cm flux density $\ge 1.5$ Jy (2 Jy for
decl. $<0 ^\circ$) at any epoch between 1994 and 2003.
There are 133 AGN that satisfy these criteria, $\sim 95$\% of
which are flat-spectrum blazars.  Superluminal speeds have been
obtained for nearly the entire sample, and the speed distribution is
consistent with a power-law distribution of bulk Lorentz factors in the
parent population (Lister et al., in preparation).

%\begin{itemize}
%\item J2000 Declination $>$ $-$20$^\circ$
%\item Galactic latitude $|b| > 2.5^\circ$
%\item VLBA 2 cm flux density $\ge 1.5$ Jy (2 Jy for decl. $<0 ^\circ$) at any epoch between 1994 and 2003.
%\end{itemize}

%In this paper we
%describe Monte Carlo simulations of AGN samples selected on beamed jet
%flux density, which we use to explore the demographics of blazars that
%will likely be seen by GLAST. We describe an extension of the
%MOJAVE program through the GLAST mission that will provide useful
%parameters on blazar jets in complete regions of the radio/gamma-ray
%flux density plane.

In principle, the beamed flux density from an AGN jet should be a
function of the bulk Lorentz factor of the emitting region, its
intrinsic (un-beamed) luminosity, the observer viewing angle, and the
redshift. Under the simplifying assumptions of a non-varying Lorentz
factor, straight jets, and flat radio spectral index, the probability
density functions for these four quantities can be derived in
connection with Monte Carlo techniques to produce simulated
flux-limited AGN samples (e.g.,
\citep{LM97}). If we can assume an intrinsic scaling law between the
un-beamed luminosities of jets in the radio and gamma-ray regimes, it
is possible to produce a simulated distribution of blazars in the
radio/gamma-ray plane. For this study, we have chosen a simple linear
scaling (e.g., \citep{SSM93,SS94}) of the form $f =
\sigma{L_{\gamma-\mathrm{ray}}/ L_{\mathrm{radio}}} +
\mathrm{const.},$ where $\sigma$ is a Gaussian deviate of unit
width. We then calculate the beamed jet luminosities according to
$P_{\mathrm{radio}} = L_{\mathrm{radio}}\delta^2$, $P_{\mathrm{SSC}} =
L_{\mathrm{radio}}\delta^{2-\alpha} f$, and $P_{\mathrm{ECS}} =
L_{\mathrm{radio}}\delta^{3-2\alpha} f,$ where $\delta$ is the jet
Doppler factor and $\alpha = -1.4$ is the typical gamma-ray spectral
index \citep{SHM01}. The synchrotron-self Compton (SSC) and external
Compton scattering (ECS) models differ mainly in the source of the
seed photons for the inverse-Compton emission, and therefore have
different beaming dependencies \citep{Derm95}.

%\begin{eqnarray}
%P_{\mathrm{radio}} &=& L_{\mathrm{radio}}\delta^2 \cr
%P_{\mathrm{SSC}} &=& L_{\mathrm{radio}}\delta^{2-\alpha} f \cr
%P_{\mathrm{ECS}}& =& L_{\mathrm{radio}}\delta^{3-2\alpha} f, 
%\end{eqnarray}
Fig. 1 shows the predicted distributions in the
gamma-ray/radio flux density plane for these two emission
models. In these simulations we use the fitted intrinsic luminosity
parameters of \cite{CL07}, and assume a parent Lorentz factor
distribution of the form $n(\Gamma)d\Gamma \propto \Gamma^{-1.5}
d\Gamma$, with $\Gamma_{\mathrm{max}} = 30$. The latter limit, which
comes from the highest speeds observed in radio-loud blazar jets
\citep{KL04}, creates an upper envelope in Figure 1 that corresponds to
sources with the maximum possible Doppler factors ($\delta \simeq
60$). 

These simulations indicate that despite a relatively small scatter in
the intrinsic radio/gamma-ray luminosity ratio, a very large range of
gamma-ray flux density is expected at a given radio flux density
level. This is entirely due to the large range of possible Doppler
factor values in the population, and the larger amount of Doppler
boosting in the gamma-ray regime. This implies that if a population of
intrinsically radio bright yet gamma-ray weak blazars exists, its
signal will likely be wiped out by the large gamma-ray flux scatter
associated with Doppler beaming. Although the SSC model predicts a
slightly lower median gamma-ray flux for bright radio-loud blazars, it
will be difficult to distinguish between these two emission models
without additional jet Doppler factor information. 

\section{The Extended MOJAVE Survey}

The MOJAVE survey will continue to provide publicly available
high-resolution VLBA data on bright blazars throughout the GLAST
mission. The sample has recently been expanded to include all known
high confidence EGRET blazars above declination $-20^\circ$, as well
as 33 nearby, low-luminosity AGN. Following the launch of GLAST, we
will also add up to 100 of the brightest gamma-ray AGN above
declination $-30^\circ$ which have sufficient flux density for direct
VLBA fringe detection ($\sim 200$ mJy at 2 cm).  This will provide
complete kinematic information in regions A through D in Figure
1. Each individual AGN will be observed at a cadence appropriate to
its angular expansion rate, such that individual jet features can be
reliably identified across the epochs. These cadences range from once
every two months for fast sources like BL Lac, 3C 273, and 3C 120, to
once every two to three years for the slowest-varying jets. Additional
programs at the U. Michigan, Owens Valley, and Effelsberg radio
observatories will also provide dense flux-density monitoring data at
cm and mm wavelengths on the entire sample.

%\begin{theacknowledgments}
\medskip
The author wishes to acknowledge the contributions of M.Cara and the
other members of the MOJAVE team. 
%H. and M. Aller, T. Arshakian, M. Cara, M. Cohen, D. Homan, M. Kadler,
%K. Kellermann, Y. Kovalev, A. Lobanov, E. Ros, R. Vermeulen, \&
%J. A. Zensus. 
The MOJAVE project is supported under National Science
Foundation grant 0406923-AST.
%\end{theacknowledgments}

\begin{figure}[t]
\caption{Simulated blazar populations based on the MOJAVE radio quasar
sample. Plotted on the x-axis is compact radio jet (VLBA) luminosity
at 15 GHz. The y-axis represents the predicted gamma-ray flux density,
assuming a simple linear scaling between intrinsic radio and gamma-ray
jet luminosity with Gaussian scatter. The left panel uses an external
synchrotron seed photon model to calculate the beamed gamma-ray
emission, while the right panel assumes a synchrotron self-Compton
model.  }

\includegraphics[scale=0.4]{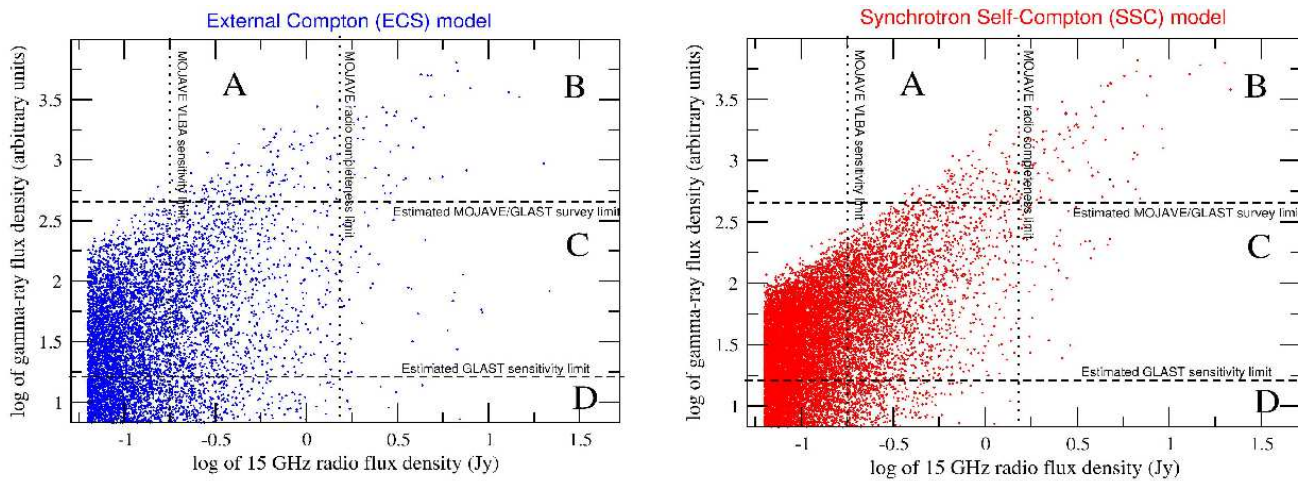}
\end{figure}

%\begin{figure}
%  \resizebox{6pc}{!}{\includegraphics{fig1.eps}}
%  \resizebox{6pc}{!}{\reflectbox{\includegraphics{escher}}}
%\caption{The caption}
%\end{figure}

% choose bibtex style depending on layout style and options used in
% sample:

%\doingARLO[\bibliographystyle{aipproc}]
%          {\ifthenelse{\equal{\AIPcitestyleselect}{num}}
%             {\bibliographystyle{arlonum}}
%             {\bibliographystyle{arlobib}}
%          }
%\bibliography{sample}

\begin{thebibliography}{}

\bibitem[Cara \& Lister(2007)]{CL07} Cara, M. \& Lister, M. 2007, astro-ph/0702449

\bibitem[Dermer(1995)]{Derm95} Dermer, C.~D.\ 1995, ApJl, 
446, L63 
\bibitem[Kellermann et al.(2004)]{KL04} Kellermann et al., 2004, ApJ,
609, 539

%\bibitem[Lister(2003)]{L03} Lister, M.~L.\ 2003, ApJ, 599, 105 

\bibitem[Lister \& Homan(2005)]{LH05} Lister, M.~L., \& 
Homan, D.~C.\ 2005, AJ, 130, 1389 

\bibitem[Lister \& Marscher(1997)]{LM97} Lister, M.~L.~\&
Marscher, A.~P.\ 1997, ApJ, 476, 572

\bibitem[Salamon \& Stecker(1994)]{SS94} Salamon, M.~H., \& Stecker,
F.~W.\ 1994, ApJ, 430, L21 

\bibitem[Sreekumar et al.(2001)]{SHM01} Sreekumar, P., 
Hartman, R.~C., Mukherjee, R., \& Pohl, M.\ 2001, AIP Conf.~Proc., 587, 314 

\bibitem[Stecker et al.(1993)]{SSM93} Stecker, F.~W., Salamon, M.~H., \& Malkan, M.~A.\ 1993, ApJ, 410, L71 

\end{thebibliography}

\end{document}